# Qualitative difference between the angular anisotropy parameters in fast electron scattering and photoionization

**M. Ya. Amusia*+, L. V. Chernysheva+, and E. Z. Liverts***

*Racah Institute of Physics, Hebrew University, 91904 Jerusalem, Israel
+A.F.Ioffe Physical-Technical Institute, 194021 St.-Petersburg, Russia

It is demonstrated for the first time that in spite of well known big similarities between atomic ionization by photons and fast electrons, a qualitative difference exists in angular anisotropy parameters of electrons knocked out in these processes. The difference is disclosed here and attributed to distinction between normal (transverse) and virtual (longitudinal) photons. Formulas are derived for dipole and non-dipole angular anisotropy parameters in fast electron-atom scattering. The ratio of quadrupole-to-dipole matrix elements is determined by the parameter $\omega R/\upsilon \ll 1$ where $\omega$ is the transferred in collision energy, $R$ is the ionized shell radius and $\upsilon$ is the speed of projectile. This factor can be much bigger than in the case of photoionization, where one has the speed of light $c$ that is much bigger than $\upsilon$.

We illustrate general formulas by concrete results for outer s-subshells of noble gas atoms Ar and Xe. Even for very small transferred momentum $q$, in the so-called optical limit, the deviation from photoionization case is prominent and instructive.

**1.** In this Letter we present for the first time the angular anisotropy parameters of secondary electrons removed of the atom in fast electron-atom collisions. We consider the distribution relative to momentum $\vec{q}$, transferred to the target in the collision process. We limit ourselves to so small $q$ that permits to consider only several lower polynomials in the angular distribution of the secondary electrons. This allows studying not only dipole, but also monopole, quadrupole and octupole transition matrix elements as function of both $\omega$ and $q$.

Non-dipole corrections to photoionization were presented for the first time long ago [1] but due to experimental difficulties were observed and investigated starting only about ten - fifteen years ago (see [2] and references therein). This permitted to study quadrupole continuous spectrum matrix elements of atomic electrons that in the absolute cross photoionization cross-section are unobservable in the shadow of much bigger dipole contribution. To study non-dipole parameters high intensity sources of continuous spectrum electromagnetic radiation were used (see [3] and references therein). In principle, however, the information from photoionization studies does not include $q$-dependences and monopole matrix elements.

By the order of magnitude the ratio quadrupole-to-dipole matrix elements in photoionization is characterized by the parameter $\omega R/c$, where $\omega$ is the photon energy, $R$ is the ionized shell radius and $c$ is the speed of light. For photon energies up to several keV that include ionization potential of the inner 1s-subshell even for medium atoms, it is $\omega R/c \ll 1$. In the absolute cross-sections dipole and quadrupole terms do not interfere, so that the ratio of quadrupole to dipole contributions in the absolute cross section is given by the second power of the parameter $\omega R/c \ll 1$ and some of these terms are canceling each other. As to the angular distribution, it includes the dipole-quadrupole interference terms in the first power of $\omega R/c \ll 1$ and therefore the relative role of quadrupole terms is bigger.

Quite long ago the fast charged particle inelastic scattering process was considered as a "synchrotron for poor" [4]. This notion reflects the fact that the fast charge particle inelastic scattering is similar to photoionization, since it is mainly determined by the dipole contribution. But contrary to the photoionization case the ratio "quadrupole-to-dipole" contributions can be much bigger, since instead of $\omega R/c \ll 1$ they are determined by $\omega R/\upsilon$, where $\upsilon$ is the speed of the projectile and $\omega$ is the transferred energy. Since $1 \ll \upsilon \ll c$, the quadrupole term in inelastic scattering is relatively much bigger[1]. The transferred in collision momentum *q* is not bound to the transferred energy $\omega$ by a relation similar to $\omega = aq$, with $a$ being a constant. Therefore the collision experiment gives an extra degree of freedom to control the atomic reaction to the transferred energy and linear moment. This stimulates the current research, the aim of which is to derive formulas for the angular anisotropy parameters of electrons emitted off the atom in its inelastic scattering with a fast charged projectile, and to perform calculations of these parameters as functions of $\omega$ and *q*.

Deep similarity between photoionization and fast electron scattering brought to a belief that not only the total cross-section, but also angular anisotropy parameters are either the same of similar. As it is shown below, this is incorrect even in the limit $q \to 0$.

In this Letter, we investigate the differential cross-section of inelastic scattering upon atom as a function of the angle $\theta$ between the momentum of the emitted in collision process electron and the direction of $\vec{q}$. As it is known, the fast charged particle inelastic scattering cross section is proportional to the so-called generalized oscillator strength (GOS) density. Thus, we study in this Letter the GOS density angular distribution as a function of $\theta$.

We go beyond the one electron Hartree-Fock approximation by including multi-electron correlations in the frame of the random phase approximation with exchange (RPAE) that was successfully applied to studies of photoionization and fast electron scattering [5].

**2.** The cross-section of the fast electron inelastic scattering upon an atom with ionization of an electron of *nl* subshell can be presented as [6]

$$\frac{d^2\sigma_{nl}}{d\omega do} = \frac{2\sqrt{(E-\omega)}}{\sqrt{E}\omega q^2}\frac{dF_{nl}(q,\omega)}{d\omega}. \tag{1}$$

Here $dF_{nl}(q,\omega)/d\omega$ is the differential in the ionized electron energy $\varepsilon = \omega - I_{nl}$ the GOS density, $I_{nl}$ is the *nl* subshell ionization potential.

In one-electron approximation the GOS density differential both in the emission angle and energy of the ionized electron with linear momentum $\vec{k}$ from a subshell with principal quantum number *n* and angular momentum *l* is given by the following formula:

$$\frac{df_{nl}(q,\omega)}{d\Omega} = \frac{1}{2l+1}\frac{2\omega}{q^2}\sum_m \left|\left\langle nlms\right|\exp(i\vec{q}\vec{r})\left|\varepsilon\vec{k}s\right\rangle\right|^2. \tag{2}$$

Here $\vec{q} = \vec{p} - \vec{p}\,'$, with $\vec{p}$ and $\vec{p}\,'$ being the linear moments of the fast incoming and outgoing electrons determined by the initial $E$ and final $E'$ energies as $p = \sqrt{2E}$ and $p' = \sqrt{2E'}$, $\Omega$ is the solid angle of the emitted electron, *m* is the angular momentum projection, *s* is the electron spin. Note that $\omega = E - E'$ and $\varepsilon = \omega - I_{nl}$ is the outgoing electron energy.

The values of $\omega$ are limited by the relation $0 \le \omega \le pq$, contrary to $\omega = cq$ for the case of photoeffect. In order to consider the projectile as fast, its speed must be much higher than the speed of electrons in the ionized subshell, i.e. $\sqrt{2E} \gg R^{-1}$. The transferred to the atom momentum *q* is considered as small if $qR \le 1$.

---

[1] Atomic system of units is used in this paper: electron charge *e*, its mass *m* and Plank constant $\hbar$ being equal to 1, $e = m = \hbar = 1$

Expanding $\exp(i\vec{q}\vec{r})$ into a sum of products of radial and angular parts and performing analytic integration over the angular variables, one obtains for GOS in one-electron Hartree-Fock approximation:

$$g_{nl,kl',L}(q) \equiv \int_0^\infty R_{nl}(r)\, j_L(qr) R_{kl'}(r) r^2 dr\,, \tag{3}$$

where $j_L(qr)$ are the spherical Bessel functions.

We suggest measuring the angular distribution of the emitted electrons relative to $\vec{q}$. It means that the z-axis coincides with the direction of $\vec{q}$ and hence one has to put $\theta_{\vec{q}} = \varphi_{\vec{q}} = 0$ in Eq. (2). Since we have in mind ionization of a particular *nl* subshell, for simplicity of notation and due to energy conservation in the fast electron inelastic scattering process leading to $k = \sqrt{2(\omega - I_{nl})}$, let us introduce the following abbreviations $g_{nl,kl',L}(q) \equiv g_{kl'L}(q)$.

The GOS formulas can be generalized in order to include inter-electron correlations in the frame of RPAE. This is achieved by substituting $g_{kl'L'}(q)$ by modulus $\tilde{G}_{kl'L'}(q)$ and the scattering phases $\delta_{l'}$ by $\bar{\delta}_{l'} = \delta_{l'} + \Delta_{l'}$, where the expressions $G_{kl'L'}(q) \equiv \tilde{G}_{kl'L'}(q)\exp(i\Delta_{l'})$ are solutions of the RPAE set of equations [7]:

$$\begin{aligned} &\left\langle \varepsilon l' | G^L(\omega,q) | nl \right\rangle = \left\langle \varepsilon l' | j_L(qr) | nl \right\rangle + \\ &+\left( \sum_{\varepsilon'' l'' \le F, \varepsilon''' l''' > F} - \sum_{\varepsilon'' l'' < F, \varepsilon''' l''' \le F} \right) \frac{\left\langle \varepsilon''' l''' | G^L(\omega,q) | \varepsilon'' l'' \right\rangle \left\langle \varepsilon'' l'', \varepsilon l' | U | \varepsilon''' l''', nl \right\rangle_L}{\omega - \varepsilon_{\varepsilon''' l'''} + \varepsilon_{\varepsilon'' l''} + i\eta(1 - 2n_{\varepsilon''' l'''})} \cdot \end{aligned} \tag{4}$$

Here $\le F (> F)$ denotes summation over occupied (vacant) atomic levels in the target atom. Summation over vacant levels includes integration over continuous spectrum, $n_{\varepsilon l}$ is the Fermi step function that is equal to 1 for $nl \le F$ and 0 for $nl > F$; the Coulomb interelectron interaction matrix element is defined as $\left\langle \varepsilon'' l'', \varepsilon l' | U | \varepsilon''' l''', nl \right\rangle_L = \left\langle \varepsilon'' l'', \varepsilon l' | r_<^L / r_>^{L+1} | \varepsilon''' l''', nl \right\rangle$ - $\left\langle \varepsilon'' l'', \varepsilon l' | r_<^L / r_>^{L+1} | nl, \varepsilon''' l''' \right\rangle$. In the latter formula notation of smaller (bigger) radiuses of $r_<(r_>)$ of interacting electron coordinates comes from the well-known expansion of the Coulomb interelectron interaction. The necessary details about solving (4) one can find in [8].

For differential in the outgoing electron angle GOS density of *nl* subshell $dF_{nl}(q,\omega)/d\Omega$ the following relation are valid in RPAE

$$\begin{aligned} &\frac{dF_{nl}(q,\omega)}{d\Omega} = \sum_{L'L''} \frac{dF_{nl}^{L',L''}(q,\omega)}{d\Omega} = \frac{\omega\pi}{q^2} \sum_{L'L''} (2L'+1)(2L''+1) i^{L'-L''} \times \\ &\sum_{l'=|L'-l|}^{L'+l} \sum_{l''=|L''-l|}^{L''+l} \tilde{G}_{kl'L'}(q) \tilde{G}_{kl''L''}(q) i^{l''-l'} (2l'+1)(2l''+1) e^{i(\bar{\delta}_{l'} - \bar{\delta}_{l''})} \begin{pmatrix} L' & l & l' \\ 0 & 0 & 0 \end{pmatrix} \begin{pmatrix} l'' & l & L'' \\ 0 & 0 & 0 \end{pmatrix} \\ &\sum_{L=|l'-l''|}^{l'+l''} P_L(\cos\theta)(-1)^{L+l}(2L+1) \begin{pmatrix} l' & L & l'' \\ 0 & 0 & 0 \end{pmatrix} \begin{pmatrix} L & L' & L'' \\ 0 & 0 & 0 \end{pmatrix} \begin{Bmatrix} L & L' & L'' \\ l & l'' & l' \end{Bmatrix}. \end{aligned} \tag{5}$$

The partial value of GOS $F_{nl}(q,\omega)$ in RPAE is obtained from (5) by integrating over $d\Omega$, leading to the following expressions:

$$F_{nl}(q,\omega)=\sum_{L'}F_{nl}^{L'}(q,\omega)=\frac{4\omega\pi^2}{q^2}\sum_{L'}(2L'+1)\sum_{l'=|L'-l|}^{L'+l}[\tilde{G}_{kl'L'}(q)]^2(2l'+1)\begin{pmatrix} L' & l & l' \\ 0 & 0 & 0 \end{pmatrix}^2 \quad . \tag{6}$$

Note that at small $q$ the dipole contribution in GOSes $F_{nl}(q,\omega)$ dominates and is simply proportional to the photoionization cross-section $\sigma_{nl}(\omega)$ [5]. To compare the results obtained with known formulas for the photoionization with lowest order non-dipole corrections taken into account, let us consider so small $q$ that it is enough to take into account terms with $L',L''\le 2$. In this case, GOS angular distribution (5) can be presented similar to the photoionization case as

$$\frac{dF_{nl}(q,\omega)}{d\Omega}=\frac{F_{nl}(q,\omega)}{4\pi}\left\{1-\frac{\beta_{nl}^{(in)}(\omega,q)}{2}P_2(\cos\theta)+q\left[\gamma_{nl}^{(in)}(\omega,q)P_1(\cos\theta)+\eta_{nl}^{(in)}(\omega,q)P_3(\cos\theta)+\right.\right.$$
$$\left.\left.\varsigma_{nl}^{(in)}(\omega,q)P_4(\cos\theta)\right]\right\}. \tag{7}$$

The obvious difference is the $q$ dependence of the coefficients and an extra term $\varsigma_{nl}^{(in)}(\omega,q)P_4(\cos\theta)$. Even in this case expressions for $\beta_{nl}^{(in)}(\omega,q)$, $\gamma_{nl}^{(in)}(\omega,q)$, $\eta_{nl}^{(in)}(\omega,q)$, and $\varsigma_{nl}^{(in)}(\omega,q)$ via $g_{kl'L'}(q)$ are too complex as compared to relations for $\beta_{nl}(\omega)$, $\gamma_{nl}(\omega)$, and $\eta_{nl}(\omega)$ in photoionization. Therefore, it is more convenient to present the results for *s*, *p*, and *d* subshells separately. We demonstrate that while $F_{nl}(q,\omega)\sim\sigma(\omega)$, similar relations are not valid for the anisotropy parameters.

Here we concentrate on *s*-subshells and compare the result obtained in the small $q$ limit with the known formula for photoionization of an atom by non-polarized light. To do this, we have to use the lowest order terms of the first three spherical Bessel functions:

$$j_0(qr)\cong 1-\frac{(qr)^2}{6};\;\; j_1(qr)\cong\frac{qr}{3}\left(1-\frac{(qr)^2}{10}\right);\; j_2(qr)\cong\frac{(qr)^2}{15}\left(1-\frac{(qr)^2}{14}\right). \tag{8}$$

The lowest in powers of $q$ term is $\tilde{G}_{11}\sim q\ll 1$[2]. Correction to $\tilde{G}_{11}$ is proportional to $q^3$. As to $\tilde{G}_{00}$ and $\tilde{G}_{22}$, they are proportional to $q^2$ with corrections of the order of $q^4$. Retaining in (7) terms of the order of $q^2$ and bigger, one has the following expression:

$$\frac{dF_{n0}(q,\omega)}{d\Omega}=\frac{F_{n0}(q,\omega)}{4\pi}\left\{1+2P_2(\cos\theta)+\frac{2}{\tilde{G}_{11}}\left[\tilde{G}_{00}\cos(\bar{\delta}_0-\bar{\delta}_1)+2\tilde{G}_{22}\cos(\bar{\delta}_1-\bar{\delta}_2)\right]P_1(\cos\theta)+\right.$$
$$\left.\frac{6\tilde{G}_{22}}{\tilde{G}_{11}}\cos(\bar{\delta}_1-\bar{\delta}_2)P_3(\cos\theta)\right\}\equiv \tag{9}$$
$$\equiv\frac{F_{n0}(q,\omega)}{4\pi}\left\{1+2P_2(\cos\theta)+q\gamma_{n0}^{(in)}(q,\omega)P_1(\cos\theta)+q\eta_{n0}^{(in)}(q,\omega)P_3(\cos\theta)\right\}$$

One should compare this relation with the similar one for photoionization of *n*0 subshell that [1]:

---

[2] As is seen from (8), we have in mind such values of $q$ that it is $qR_{nl}<1$, where $R_{nl}$ is the radius of the ionized subshell.

$$\frac{d\sigma_{n0}(\omega)}{d\Omega}=\frac{\sigma_{n0}(\omega)}{4\pi}\left\{1-P_2(\cos\theta)+\kappa\frac{6\tilde{Q}_2}{5\tilde{D}_1}\cos(\bar{\delta}_1-\bar{\delta}_2)\left[P_1(\cos\theta)-P_3(\cos\theta)\right]\right\}\equiv$$
$$\equiv\frac{\sigma_{n0}(\omega)}{4\pi}\left\{1-P_2(\cos\theta)+\kappa\gamma_{n0}(\omega)P_1(\cos\theta)+\kappa\eta_{n0}(\omega)P_3(\cos\theta)\right\}. \qquad (10)$$

where $\gamma_{n0}(\omega)=-\eta_{n0}(\omega)=\frac{6\tilde{Q}_2}{5\tilde{D}_1}\cos(\bar{\delta}_1-\bar{\delta}_2)$.

The difference between (9) and (10) is seen in the sign and magnitude of the dipole parameters and in different expressions for the non-dipole terms. This difference exists and is essential even in the so-called optical limit $q\to 0$. According to (8), there are simple relations in the $q\to 0$ limit between dipole $\tilde{D}_1$ and quadrupole $\tilde{Q}_2$ matrix elements and functions $\tilde{G}_{11}$, $\tilde{G}_{22}$: $\tilde{G}_{11}=q\tilde{D}_1/3$ and $\tilde{G}_{22}=2q^2\tilde{Q}_2/15$. With the help of relations $\tilde{G}_{00}=-q^2\tilde{Q}_2/3=-(5/2)\tilde{G}_{22}$, (9) is transformed into the following expression:

$$\frac{dF_{n0}(q,\omega)}{d\Omega}=$$
$$\frac{F_{n0}(q,\omega)}{4\pi}\left\{1+2P_2(\cos\theta)+q\frac{2\tilde{Q}_2}{\tilde{D}_1}\left[\frac{4}{5}\cos(\bar{\delta}_1-\bar{\delta}_2)-\cos(\bar{\delta}_0-\bar{\delta}_1)\right]P_1(\cos\theta)+2q\gamma_{n0}(\omega)P_3(\cos\theta)\right\} \qquad (11)$$

The deviation from (10) is evident, since the angular distribution is not expressed via a single non-dipole parameter $\gamma_{n0}(\omega)$ - a new phase difference $\bar{\delta}_0-\bar{\delta}_1$ appears. As a result, the following relations have to be valid at very small q:

$$\gamma_{n0}^{(in)}(\omega)=\frac{2\tilde{Q}_2}{\tilde{D}_1}\left[\frac{4}{5}\cos(\bar{\delta}_1-\bar{\delta}_2)-\cos(\bar{\delta}_0-\bar{\delta}_1)\right],$$
$$\eta_{n0}^{(in)}(\omega)=2\gamma_{n0}(\omega)=\frac{12}{5}\frac{\tilde{Q}_2}{\tilde{D}_1}\cos(\bar{\delta}_1-\bar{\delta}_2). \qquad (12)$$

We see that the investigation of inelastic scattering even at $q\to 0$ permits to obtain an additional characteristic of the ionization process, namely, its s-wave phase.

For $l>0$ even at very small q the relation between non-dipole parameters in photoionization and inelastic fast electron scattering are rather complex.

The similarity of general structure and considerable difference between (9) and (10) is evident. Indeed, the contribution of the non-dipole parameters can be enhanced, since the condition $\omega/c\ll q\ll R^{-1}$ is easy to achieve. Let us note that even while neglecting the terms with $q$, (10) and (11) remain different: in photoionization the angular distribution is proportional to $\sin^2\theta$ (see (10)), whereas in inelastic scattering it is proportional to $\cos^2\theta$ (see (11)). The reason for this difference is clear. In photoabsorption the atomic electron is “pushed” off the atom by the electric field of the photon, which is perpendicular to the direction of the light beam. In inelastic scattering the push acts along momentum $\vec{q}$, so the preferential emission of the electrons takes place along the $\vec{q}$ direction, so the maximum is at $\theta=0$. Similar reason explains the difference in the non-dipole terms. Note that the last term due to monopole transition in (11) is absent in photoabsorption angular distribution (10). It confirms that the angular distribution of the GOS densities is richer than that of photoionization.

**3.** In order to obtain $dF_{n0}(q,\omega)/d\Omega$ from experiment, one has to measure the yield of electrons emitted at a given angle $\theta$ with energy $\varepsilon = k^2/2 = \omega - I_{nl}$ in coincidence with the fast outdoing particle that looses energy $\omega$ and transfers to the target atom momentum $\vec{q}$. Note that $\beta_{n0}^{(in)}$ is (-4) that differs by sign and value from photoionization value $\beta_{n0} = 2$.

To calculate $dF_{n0}(q,\omega)/d\Omega$ we used the numeric procedures described at length in [8]. Calculations are performed in the frame of Hartree-Fock and RPAE approximations. As concrete objects, we choose subvalent $3s^2$ and $5s^2$ subshells of Ar and Xe. These objects are representative, demonstrating strong influence of the electron correlations both for p- and s-electrons.

Calculations are performed using equations (5-7, 9, 10) in HF and RPAE, for $q = 0.0,\ 0.1,\ 1.1,$ and $I_{3s} < \omega < I_{3s} + 5Ry$. The results for differential in emission angle GOSes $dF_{n0}(q,\omega)/d\Omega$ and non-dipole angular anisotropy parameters are presented in Fig.1-3. The GOSes are given at the so-called magic angle determined by relation $P_2(\cos\theta_m) = 0$. At $\theta_m$ the biggest, dipole contribution is zero, so the non-dipole corrections are most prominent. The lowest value of $q$ corresponds to the photoionization limit, since $qR \ll 1$ and in the considered frequency range $\omega/c < 0.05 < q_{\min} = 0.1$. The last inequality shows that we consider non-dipole corrections to the GOSes that are much bigger than the non-dipole corrections to photoionization.

As it is seen from Fig. 1, GOSes in Ar and Xe are strongly affected by electron correlations and prominently change with increase of q. Note that in Ar a new minimum appears at about 5 Ry, while in Xe the minimum near threshold disappears. The effect of 4d is profound at both q values.

Fig. 2 and 3 shows considerable difference in non-dipole parameters even with small increase of q from 0 to 0.1, and big deviation from photoionization values, even qualitative, for $\eta$ .

**4.** We performed calculations for transitions from $s$ subshells in Ar and Xe. The results demonstrate that the angular anisotropy parameters are complex and informative functions with a number of prominent variations. They depend strongly upon the outgoing electron energy and the linear momentum $q$ transferred to the atom in fast electron inelastic scattering, being strongly affected by electron correlations.

Particular attention deserves the $q \to 0$ limit. It is seen that different, by sign and value, are the dipole angular anisotropy parameters. The non-dipole parameters in their turn deviate even qualitatively from their respective photoionization values. It is amazing that in the non-relativistic domain of energies at first glance inessential difference between a virtual and real photon leads to so powerful consequences. The information that could come from studies of angular distribution of secondary electrons at small $q$ is of great interest and value. Thus, the suggested here experimental studies are desirable.

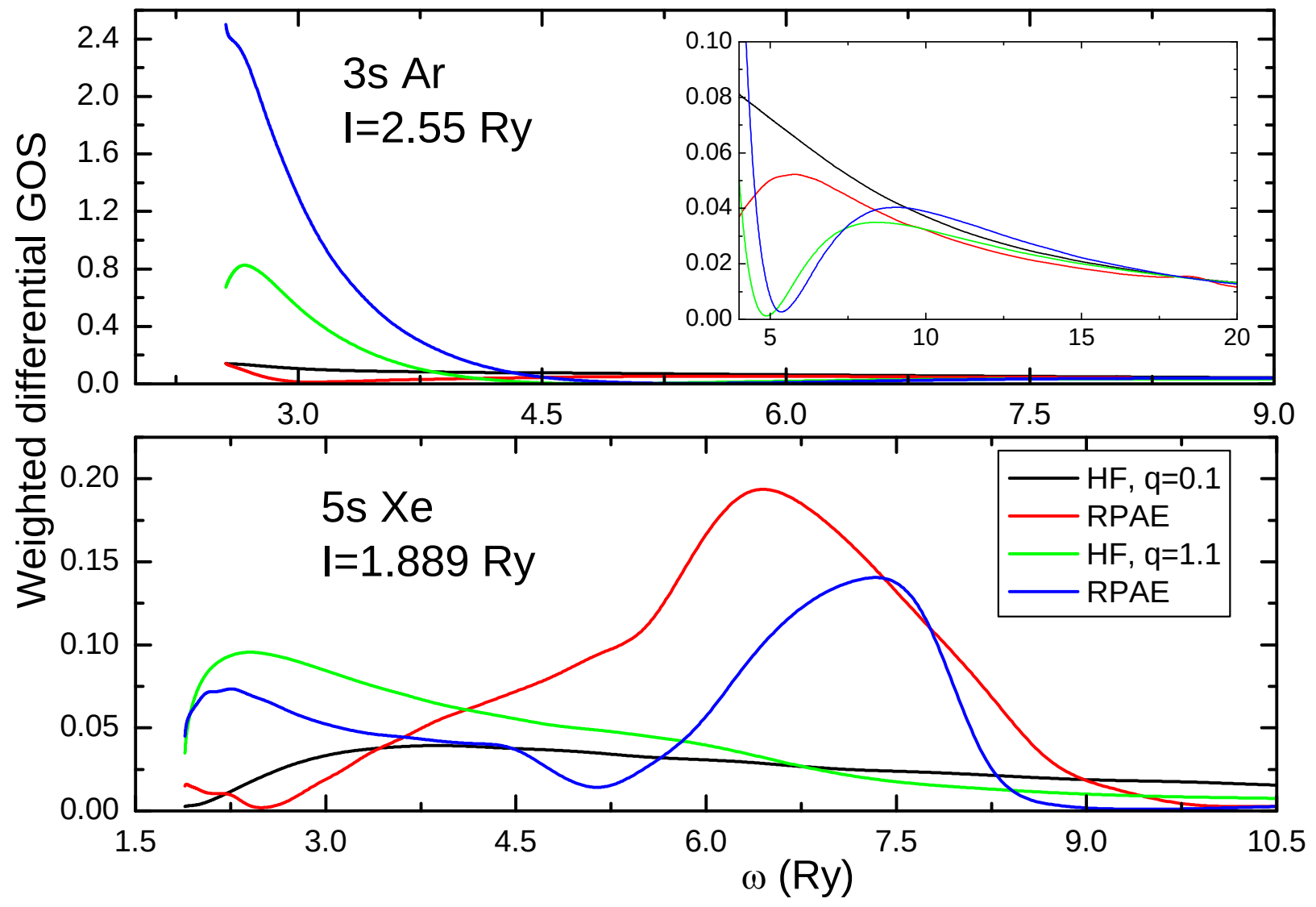

Fig. 1. Differential generalized oscillator strength (5) at magic angle $P_2(\cos\theta_m)=0$, $\theta_m$=54.736$^0$ of 3s- and 5s subshells for Ar and Xe at q=0.1, 1.1 in HF and RPAE.

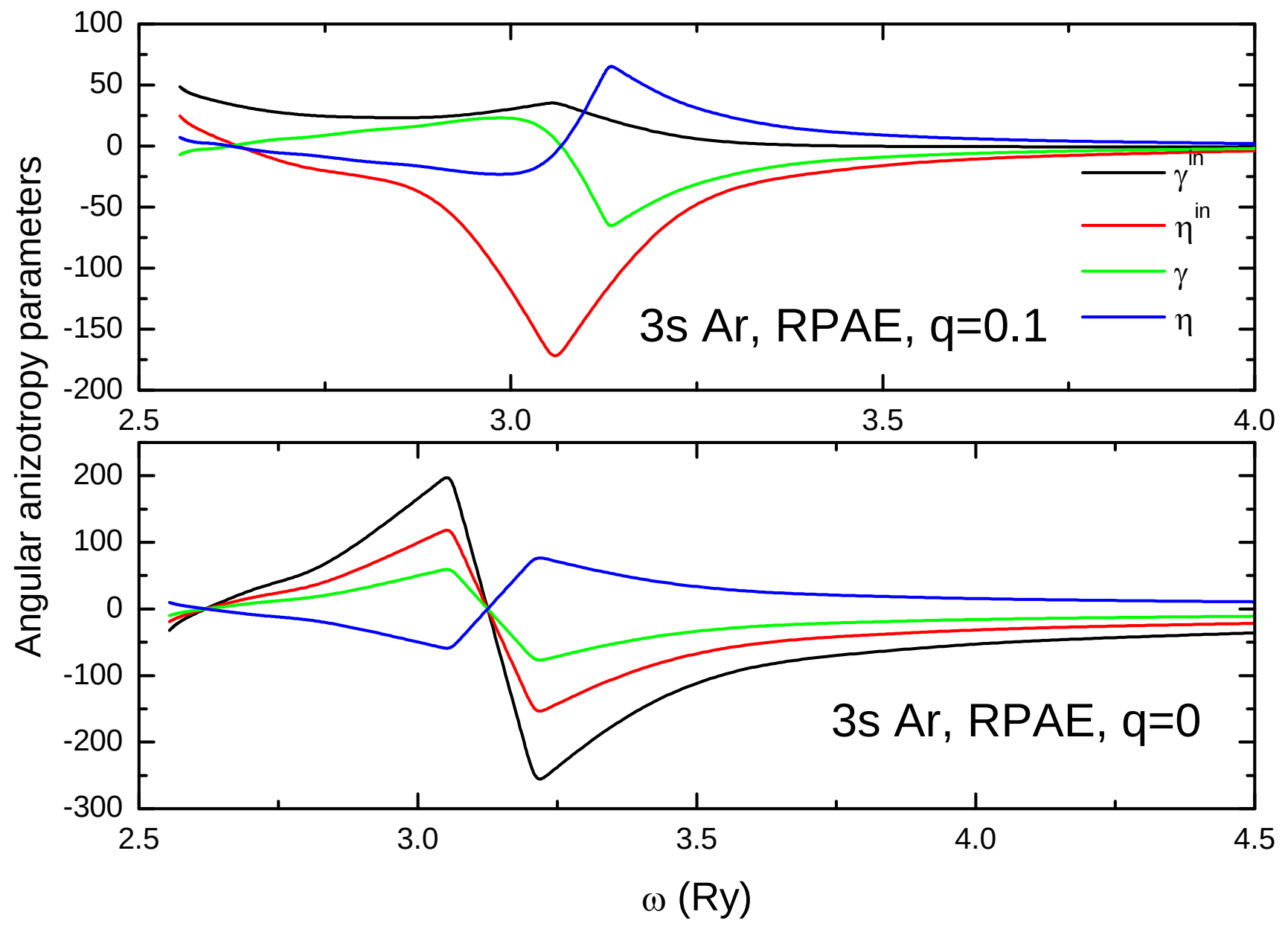

Fig. 2. Angular anisotropy non-dipole parameters of knocked-out electrons $\gamma_{3s}^{(in)}(\omega)$ and $\eta_{3s}^{(in)}(\omega)$ given by (7) and (11) at q=0.1 and q=0, compared to similar parameters in photoionization $\gamma_{3s}(\omega)$ and $\eta_{3s}(\omega)$ (10) for 3s subshell of Ar in RPAE.

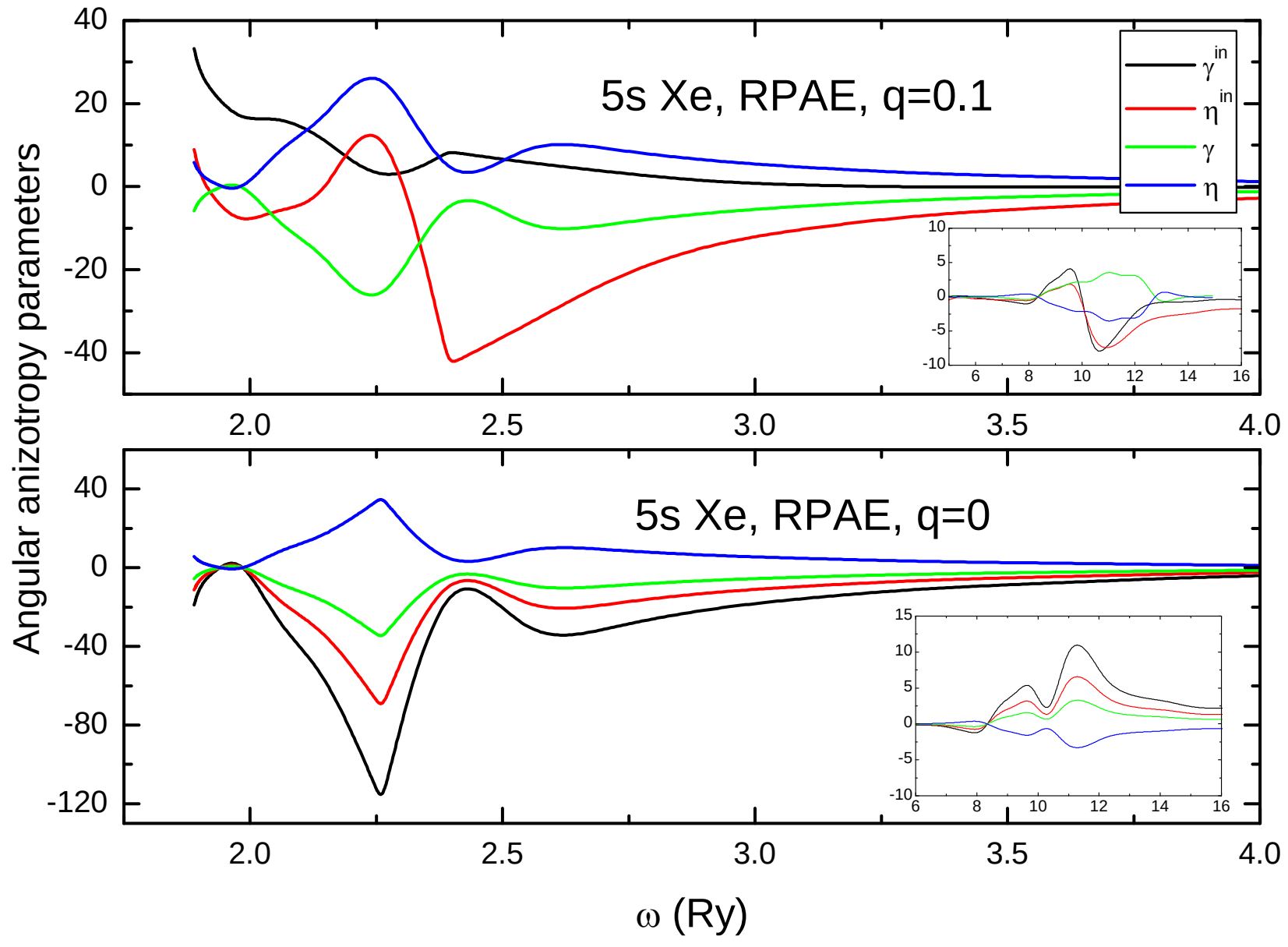

Fig. 3. Angular anisotropy non-dipole parameters of knocked-out electrons $\gamma_{5s}^{(in)}(\omega)$ and $\eta_{5s}^{(in)}(\omega)$ given by (7) and (11) at q=0.1 and q=0, compared to similar parameters in photoionization $\gamma_{5s}(\omega)$ and $\eta_{5s}(\omega)$ (10) for 5s subshell of Xe in RPAE.